\documentclass{aa}  

\usepackage{graphicx}
\usepackage{txfonts}
\usepackage{newtxtext,newtxmath} 
\usepackage[T1]{fontenc}
\usepackage{ae,aecompl}
\usepackage{amsmath}	
\usepackage{amssymb}	
\usepackage{booktabs}
\usepackage{color}
\usepackage{seqsplit}
\usepackage{ulem}
\usepackage{hyperref}
\usepackage{appendix}
\usepackage[figuresright]{rotating}
\usepackage{graphicx}
\graphicspath{{plots/}}

\newcommand{\orcid}[1]{\href{https://orcid.org/#1}{\protect\includegraphics[width=8pt]{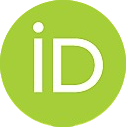}}}

\newcommand{\kms}{km\,s$^{-1}$}
\newcommand{\nstar}{N_{\rm s}}
\newcommand{\npl}{N_{\rm p}}

\newcommand{\nb}{\texttt{NBODY6++GPU}}
\newcommand{\nbm}{\texttt{NBODY6++GPU-MASSLESS}}

\newcommand{\msun}{\,{\rm M}_\odot}

\newcommand{\MLP}{\texttt{MLP}}
\newcommand{\halfmass}{r_{\rm hm}}
\newcommand{\clustermass}{M_{\rm cl}}
\newcommand{\tcross}{t_{\rm cr}}
\newcommand{\trelax}{t_{\rm rlx}}

\def\apjl{ApJ}          
\def\apjs{ApJS}          
\def\ao{Appl.~Opt.}          
\def\aap{A\&A}          
{}
{}
{}
{}

\begin{document}

\title{Dynamical evolution of massless particles in star clusters with \nbm{}}
   \titlerunning{Dynamical evolution of \MLP{}s}
   \authorrunning{Flammini Dotti et al}
\subtitle{I. Free-floating \MLP{}s}

\author{Francesco Flammini Dotti\thanks{fmfd@uni-heidelberg.de}\inst{1}\orcid{0000-0002-8881-3078},
M.B.N. Kouwenhoven\inst{2}\orcid{0000-0002-1805-0570},
Peter Berczik\inst{3,4,5}\orcid{0000-0003-4176-152X},
Qi Shu\inst{6,7}
\and
Rainer Spurzem\inst{1,6,8}\orcid{0000-0003-2264-7203}
  }

  \institute{Astronomisches Rechen-Institut, Zentrum f\"ur Astronomie, University of Heidelberg, M\"onchhofstrasse 12--14, 69120, Heidelberg, Germany
  \and
  Department of Physics, School of Mathematics and Physics, Xi'an Jiaotong-Liverpool University, 111 Ren'ai Road, \\
  Suzhou Dushu Lake Science and Education Innovation District, Suzhou Industrial Park, Suzhou 215123, P.R. China
  \and
  Nicolaus Copernicus Astronomical Centre Polish Academy of Sciences, ul. Bartycka 18, 00-716 Warsaw, Poland
  \and
  Konkoly Observatory, Research Centre for Astronomy and Earth Sciences, HUN-REN CSFK, MTA Centre of Excellence, Konkoly Thege Mikl\'os \'ut 15-17, 1121 Budapest, Hungary
  \and
  Main Astronomical Observatory, National Academy of Sciences of Ukraine, 27 Akademika Zabolotnoho St, 03143 Kyiv, Ukraine
  \and
  Kavli Institute for Astronomy and Astrophysics, Peking University, Yiheyuan Lu 5, Haidian Qu, 100871, Beijing, China
  \and
  Department of Astronomy, School of Physics, Peking University, Yiheyuan Lu 5, Haidian Qu, 100871, Beijing, China
  \and
  National Astronomical Observatories and Key Laboratory of Computational Astrophysics, Chinese Academy of Sciences, 20A Datun Rd., Chaoyang District, 100101, Beijing, China
  }

   \date{Received ---; accepted ---}

  \abstract
  {Low-mass bodies, such as comets, asteroids, planetesimals, and free-floating planets, are continuously injected into the intra-cluster environment after expulsion from their host planetary systems. These can be modeled as massless particles (\MLP{}s, hereafter). The dynamics of large populations of \MLP{}s, however, has yet received little attention in literature. }
   {We investigate the dynamical evolution of \MLP{} populations in star clusters, and characterize their kinematics and ejection rates.}
   {We present \nbm, a modified version of the $N$-body simulation code \nb{}, that allows fast integration of star clusters that contain large numbers of massless particles (\MLP{}s). \nbm{} contains routines specifically directed at the dynamical evolution of low-mass bodies, such as planets.}
   {Unlike stars, \MLP{}s do not participate in the mass segregation process. Instead, \MLP{}s mostly follow the gravitational potential of the star cluster, which gradually decreases over time due to stellar ejections and stellar evolution. The dynamical evolution of \MLP{}s is primarily affected by the evolution of the core of the star cluster. This is most apparent in the outer regions for clusters with higher initial densities. High escape rates of \MLP{}s are observed before the core-collapse, after which escape rates remain stable. Denser star clusters undergo a more intense core collapse, but this does not impact the dynamical evolution of \MLP{}s. The speeds of escaping stars are similar to those of escaping \MLP{}s, when disregarding the high-velocity ejections of neutron stars during the first 50~Myr.\\}
   {}


   \maketitle

%

\section{Introduction}

The vast majority of young stars are found in or near star forming regions that contain up to hundreds of thousands of stars \citep[e.g.,][]{Lada:2003aa}. Most of these star-forming regions disperse within several tens of millions of years, while others remain bound and become the open clusters and globular clusters in the Galaxy today \citep[e.g.,][]{de-Grijs:2007aa}. Isotope studies of meteorites suggest that even our own Solar system was once part of a star cluster that has long dispersed \citep{looney2006, portegieszwart2009}. Given that most stars are formed in crowded stellar environments \citep[e.g.,][]{Parker2020birth}, it is worthwhile investigating the evolution of planetary systems and their constituents in such environments. Stellar encounters with planetary systems, as well as gravitational interactions between the constituents of a planetary system, can result in the ejection of planets, moons, asteroids and comets, resulting in free-floating planetary debris in a star cluster. Hereafter, we will collectively refer to such bodies, with masses far below that of stars, as massless particles (\MLP{}s). 

Upon ejection from their planetary system, \MLP{}s typically obtain a velocity-at-infinity in the range 0.1–10~\kms. The velocity-at infinity is higher for prompt ejections (as a direct result of star–planet encounters) than for delayed ejection \citep[as a result of a planet–planet scattering; see][]{malmberg2011}. When an \MLP{} is ejected with a speed above the star cluster's local escape velocity, it will escape from the star cluster. When it is ejected from its host planetary system with a speed below the star cluster's local escape velocity, it remains part of the star cluster, and becomes a \MLP{} \citep[e.g.,][]{wukai2023,wukai2024}.
We will show in our study that the trajectories of such \MLP{}s generally follow the gravitational potential of the star cluster. In specific cases, a free-floating \MLP{}s can also be captured by another star \citep[e.g.,][]{kouwenhoven2010, malmberg2011,moeckel2011, parker2012, Perets:2012aa, Portegies:2020aa}. Such captured bodies may be common \citep[e.g.,][]{Siraj:2020aa}, and can potentially be identified through their orbital parameter distributions \citep[e.g.,][]{Siraj:2018aa}. 

Directly observing \MLP{}s is extremely challenging, but extrapolations and surveys have suggested that \MLP{}s are abundant in the Galactic disc. Initial estimation ranged from two Jupiter-mass \MLP{}s for every main-sequence star \citep[][]{sumi2011}, to $10^5$ \MLP{}s with masses between $10^{-8} \ \msun$ and $10^{-2} \ \msun$ per star \citep[][]{strigari2012}. Most observed \MLP{}s are detected with microlensing \citep[][]{mao1991, gould1992,abe2004, beaulieu2006, gaudi2012}. However, only several \MLP{}s have been unambiguously detected as such \citep[e.g. CFBDSIR 2149-0403 and PSO J318-22; see][, respectively]{delorme2012, liu2013}. Most of the other \MLP{}s have an uncertain mass estimates, which prevents a confirmation. As of today, 26 observed candidates have been detected. The comprehensive analysis by \cite{mroz2017} finds that \cite{sumi2011} greatly overestimates the abundance of free-floating Jupiter-mass planets in the Galactic field, and also point out that the findings of \cite{sumi2011} are inconsistent with the current planet formation theories \citep[][]{veras2012, ma2016}. The former findings are also inconsistent with surveys of young clusters \citep[][]{Pena:2016aa, ramirez2017, scholz2012, muzic2015}. Using a much larger sample of microlensing events, \cite{mroz2017} revealed a significantly lower abundance, of roughly one free-floating Jupiter-mass planet or wide-orbit Jupiter-mass planet for every four main-sequence star in the Galactic field. More recently, a survey on stellar associations by \cite{miretroig2023}, using the technique of microlensing, found thousands of free-floating planet candidates.

In this paper we present the code \nbm{}, which can efficiently model the dynamical evolution of large populations of \MLP{}s in star clusters. In the first application of this code, we consider the dynamical evolution of a population of free-floating \MLP{}s in star clusters. We also investigate the kinematics of \MLP{}s and its relation to the mass segregation process, changing the number density of star in our star cluster models. Our findings are applicable to any population of bodies can be approximated as massless, including comets, asteroids, planetesimals, and free-floating planets. 

This article is organised as follows. In Section~\ref{section:method} we describe the code \nbm{}, our numerical method and initial conditions. We discuss the numerical performance of our code, in comparison with \nb{} in Section~\ref{section:numericalperformance}. We then discuss the dynamical evolution of the star clusters and the free-floating \MLP{} populations in Section~\ref{section:results3}. Finally, we summarise our results and discuss the implications of our findings in Section~\ref{section:conclusions}.

\section{\nbm{} }\label{section:method}

\subsection{Theoretical background}

In addition to stars, star clusters contain a vast number of smaller bodies, such as planets, moons, asteroids, comets, and planetesimals. These particles are so abundant that it is impractical to directly include them as particles in $N$-body simulations. However, since these particles have a negligible mass, it is possible to neglect the gravitational force on the other bodies in the system. \MLP{}s experience the gravitational forces of the massive bodies in the system, but they do not exert gravitational forces on any of the other bodies. \nbm{} makes use of this property of \MLP{}s, which allows for a significant speed-up of simulations.

The wall-clock time of an $N$-body simulation is dominated by the regular and irregular force calculations \citep{1973JCoPh..12..389A}.
Consider a system of $N=N_s+N_p$ particles, where $N_s$ is the number of massive particles, such as stars, and $N_p$ is the number of \MLP{}s. The number of regular force calculations for each time step is then
\begin{equation}
    N_{F1} \propto N(N-1)/2 \approx N^2/2
    \quad ,
    \label{eq:regularforce}
\end{equation}
when $N\gg 1$. With the special treatment of \MLP{}s, the number of irregular force calculations for each regular time step is
\begin{equation}
    N_{F2} \propto N_s(N_s-1)/2 + N_p N_s \approx (N_s/2+N_p)N_s
    \quad , 
    \label{eq:irregularforce}
\end{equation}
when $N_s, N_p\gg 1$. 

This approach can lead to a significant speed-up when $N_p>N_s$. When $N_p=\alpha N_s$, then $N_{F1}/N_{F2}\approx (1+\alpha)^2/(1+2\alpha)$. When $\alpha=0$ then $N_{F1}/N_{F2}=0$, when $\alpha=1$ then $N_{F1}/N_{F2} \approx 1.3$, when $\alpha=10^2$ then $N_{F1}/N_{F2} \approx 51$, and when $\alpha =10^4$ then $N_{F1}/N_{F2} \approx 5000$. By distinguishing between massive particles and \MLP{}s in the $N$-body code, it is possible to efficiently integrate star clusters with a large number of small bodies, provided that the assumption of negligible mass is not violated.


\subsection{Implementation in \nbm{}}\label{section:numericalmethod}

We present \nbm{}, a modified version of \nb{} \citep[see][]{Aarseth:1999aa, Aarseth:2003aa, Spurzem:1999aa, Wang:2015aa, Wang:2016aa,kamlah2022a,spurzem2023}. In this updated version we have implemented a fast and accurate treatment of large numbers of \MLP{}s. Each particle in the code is assigned a flag that indicates whether it is treated as a massive particle or as an \MLP{}.

When calculating the gravitational interaction between particles in \nbm{}, the gravitational forces exerted by \MLP{}s are ignored. \MLP{}s are therefore never included in the force loops of any other particle. \MLP{}s are excluded from the regular and irregular force calculations for the stars. \MLP{}s are also excluded from being the primary component of a Kustaanheimo-Stiefel (KS) regularisation \citep{Kustaanheimo:419610} event, although they can participate as a companion. For details about the algorithms mentioned above we refer to~\cite{Aarseth:2003aa}.

For each star in the system we identify its nearest neighbour using the  \cite{1973JCoPh..12..389A} scheme.  
The candidates for KS-regularisation have to satisfy two additional conditions in \nb{} and \nbm: (i) $\vec{R}\cdot\vec{V}< 0.1 \sqrt{GMR}$ and (ii) $|\vec{a}_{\rm pert}| R^2 /GM < 0.25$. Here, $R$ is the scalar distance, $\vec{R}$ the relative position, and $\vec{V}$ the relative velocity. $\vec{a}_{\rm pert}$ is the vectorial differential acceleration exerted by other perturbing particles, $M$ is the sum of the masses of the two candidates, and $G$ is the gravitational constant. For further details, we refer to \cite{spurzem2023}.\\
From the masses, positions and velocities of the particles we subsequently derive the orbital parameters of each binary system. All receding stars and \MLP{}s located beyond twice the star cluster's tidal radius are treated as escapers \cite[see, e.g.,][]{Aarseth:2003aa}. 

Stellar evolution and binary evolution are implemented following the prescriptions of the Cambridge stellar evolution package and their improvements \citep{Eggleton:1989aa, Hurley:2000aa, Hurley:2002ab, Hurley:2005aa, Belczynski:2007aa,kamlah2022a}. We adopt level~C, as described in Table~A1 of \cite{kamlah2022a}. For relativistic kicks and mergers, we use the methodology described in \protect\cite{arcasedda2023,arcasedda2024a,arcasedda2024b}.


\section{Numerical performance}\label{section:numericalperformance}

\begin{figure}
   \includegraphics[width=1.0\columnwidth]{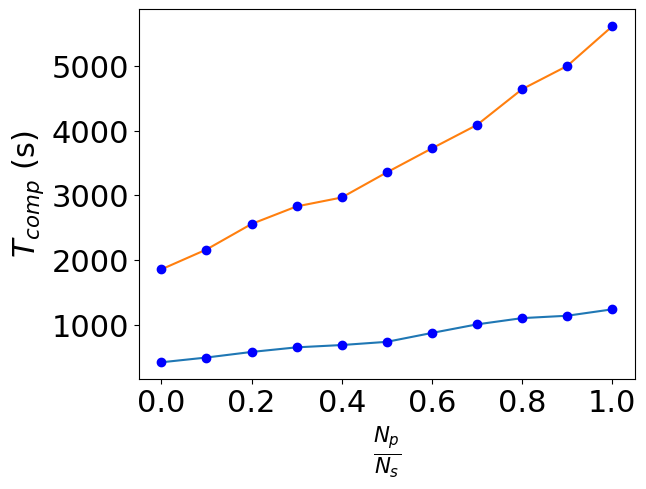}
  \caption{Wall-clock time ($T_{\rm CPU}$) consumption for carrying out star cluster simulations with different numbers of stars, $\nstar$ and different numbers of \MLP{}s, $\npl$. Each calculation is carried out for 10 $N$-body units. Simulations are performed for $\nstar=64\,000$ (blue curve) and $\nstar=128\,000$ (orange curve), and with an initial \MLP{}-to-star ratio, $\npl/\nstar$, ranging from zero to unity (blue dots).
  \label{fig:time_used}}
\end{figure}

\begin{figure}
  \includegraphics[width=1.0\columnwidth]{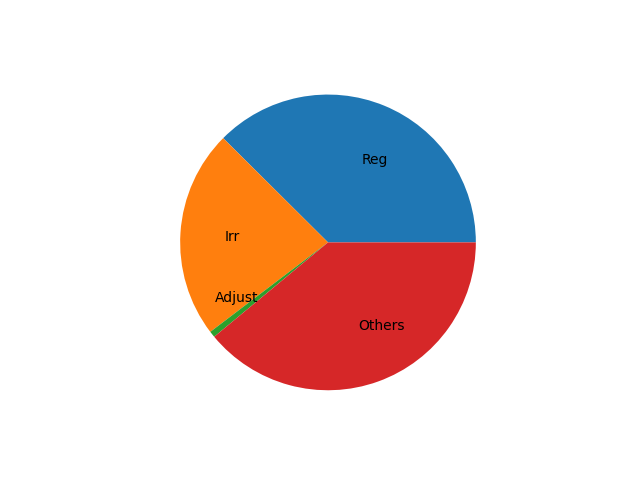}\\
   \includegraphics[width=1.0\columnwidth]{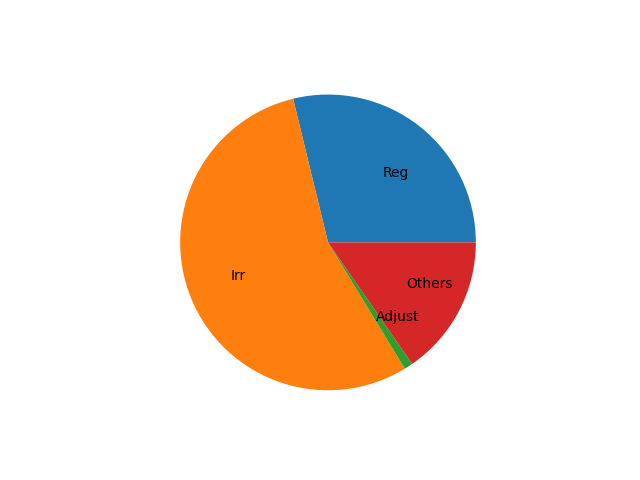}
  \caption{Wall-clock calculation times for different processes in the simulation with $\nstar = 64\,000$ and $ \npl= 64\,000  $ (top pie chart) and $\nstar = 128\,000$ and $\npl = 0$ (bottom pie chart) in Figure~\ref{fig:time_used}. The chart presents the wall-clock time for time intervals of the total simulation time, 10 code units. Different numerical processes are indicated with the colours: regular force calculations (blue), irregular force calculations (orange), adjustment processes (green), and others (red).
  It is immediately noticeable that the irregular force calculations is taking a consistent overall less time in the simulation with massless particles.
}
 \label{fig:PE}
\end{figure}

\begin{figure}[h]
\centering
   \includegraphics[width=1.0\columnwidth]{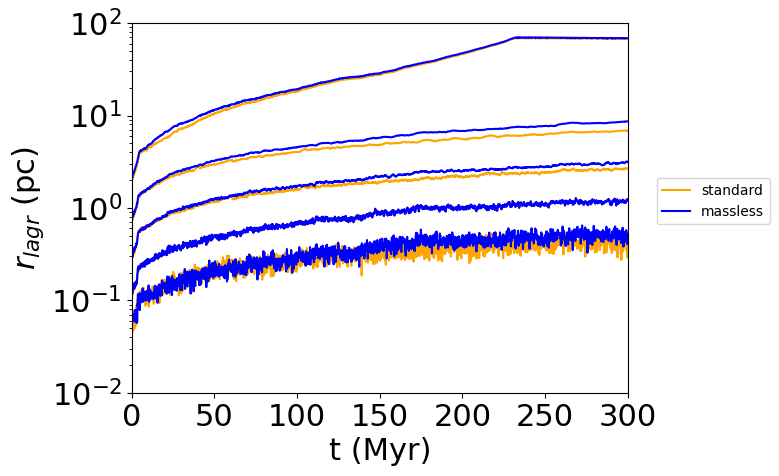}
  \caption{Comparison between the  Lagrangian radii of the \MLP{}s in the C12.8k model (which is also used for the main simulations, see next sections), using the standard version of \nb{} (\textit{in orange}) and using \nbm{} (\textit{in blue}). The curves show the 0.1 \% ({\it bottom}), 1 \%, 10\%, 50\% and 90\% ({\it top}) Lagrangian radii for both models.
  The differences between the two models is minimal, and is a consequence of the stochastic nature of the clusters and numerical noise. }
  \label{fig:standardvsmassless}
\end{figure}

The wall-clock time consumption for a system consisting of $N$ gravitationally-interacting particles scales approximately as 
\begin{equation} \label{eq:wallclock}
    \Delta T_{\rm CPU} 
    \approx k_1 N^2+ k_2 N  \overline{N}_n + k_3 + k_4 N^2 
    + \Delta T_{\rm KS} + \Delta T_{\rm comm} 
\end{equation}
Here, $k_1$ denotes the regular force computation coefficient, $k_2$ denotes the irregular (neighbour) force coefficient, and $\overline{N}_n$ is the average neighbour number, which is typically $50-200$ \citep{1973JCoPh..12..389A}. The coefficient $k_3$ represents the computational time that is (mostly) independent of $N$ and $\overline{N}_n$ \citep[for example, processes such as simulation parameter updates; see][for details]{Huang:2016aa}. The time required for diagnostics (e.g., the energy conservation checks) is governed by the coefficient $k_4$. The term $\Delta T_{\rm KS}$ is the time required for integrating two-body KS regularisation pairs \citep[e.g.,][]{Kustaanheimo:419610}, few-body chains, and hierarchical systems. Finally, $\Delta T_{\rm comm}$ represents the overhead term that accounts for parallel computation.

The contribution of the coefficients $k_3$ and $k_4$ to the wall-clock time is negligible for our simulations. As we carry out energy checks every few million years, $k_4 \ll k_1$. Moreover, $\Delta T_{\rm KS}$ is small (see also the KS implementation of \MLP{}s above) and $\Delta T_{\rm comm}$ is negligible, so that for a star cluster with $N=\nstar+\npl$ bodies, Eq.~(\ref{eq:wallclock}) thus reduces to
\begin{equation} \label{eq:wallclock2}
	\Delta T_{\rm CPU} 
    \approx k_1 N^2 + k_2 N \overline{N}_n = k_1 (\nstar^2 +2 \nstar \npl 
    + \npl^2) + k_2 (\nstar+\npl) \overline{N}_n 
    \quad .
\end{equation}
As we disable the interaction between \MLP{}s, the $N_p^2$ and $N_s^2$ terms in the force calculations can be neglected. Also, the gravitational influence from the \MLP{}s on the massive particles is disabled, so the term $\nstar \npl$ is reduced by half. With these modifications, the above expression then reduces to
\begin{equation} \label{eq:wallclock4}
    \Delta T_{\rm CPU} 
    \approx ( k_1 \nstar 
    + k_2 \overline{N}_n) \npl 
    + k_1 \nstar^2 + k_2 \nstar \overline{N}_n 
    \quad.
\end{equation}

Our test simulations were carried out with star clusters initialised with identical parameters, except for the number of stars and \MLP{}s. Each of the simulation is evolved for 10 $N$-body units. We first model a star cluster without \MLP{}s. We then add 10\% of the total number of stars for each model, until the number of \MLP{} is equal to the number of stars. We use one set of models with 64\,000 stars and one set of models with 128\,000 stars. 
We carry out our simulations on a workstation with an an Intel CPU (i9--9960X CPU @ 3.00GHz) and two NVIDIA GPU cards (GeForce RTX 2080 Ti). %

The wall-clock simulation time, $\Delta T_{\rm CPU}$, for different values of $\nstar$ and $\npl$ is shown in Figure~\ref{fig:time_used}, from stars-only simulations up to $\nstar$ = $\npl$. 

Figure~\ref{fig:PE} shows the distribution of the wall-clock time consumption over different components, for the model with 64\,000 stars and 64\,000 \MLP{}s and the model with 128\,000 stars. The distribution on the pie chart is relative to the total time of the simulation and not to the second used in the simulation. The regular force calculations, irregular force calculations, and the adjustment procedures provide the largest contributions to the wall-clock time consumption.

The main difference in the computational time of the two simulations is related to the regular and irregular forces, as the other processes are also related to hardware properties and auxiliary softwares used on the machine. The regular forces  takes 13 \% less time in the star-\MLP{} simulations. These forces are updated less frequently than the irregular forces, so the difference between the two simulations becomes more noticeable with a larger number of $N$-body temporal unit. The main contribution resides in the irregular forces, where the star and \MLP{} simulation takes only $\approx $30 \% of the time than the star-only simulation. The entire simulation, which lasts for 10~$N$-body units (NBU), has a net gain in total computational time of $\approx$ 600~s (see Figure~\ref{fig:time_used}).
Our comparison between \nb{} and \nbm{} indicates that both codes give the same performance (apart from statistical fluctuations) for stars-only clusters, but the special treatment of \MLP{}s in \nbm{} significantly speeds up the integration (depending on the number of \MLP{}s relative to the number of stars). As planets, comets, and other debris can be safely treated as \MLP{}s, the total number of such particles and their kinematic properties do not affect the dynamical evolution of the stellar population. Since \MLP{}s also do not interact with each other, all our results can be easily scaled up to star clusters with a much larger number of \MLP{}s.

Finally, in Figure~\ref{fig:standardvsmassless} we present a comparison between the standard version of \nb{} and  the code presented in this paper, \nbm{}. We carry out simulations for model C12.8k (see Section~\ref{section:initialconditions}) for 300 Myrs. Both simulations have identical initial conditions. Both simulations have the same initial conditions for all particles, using a Mercury mass for all the \MLP{}s. The only difference is that in the standard \nb{} version the \MLP{}s are included in the regular and irregular forces calculations. The difference in the Lagrangian radii is minimal, which justifies our treatment of \MLP{}s.

\section{Dynamical evolution of \MLP{} populations}\label{section:results3}


\subsection{Initial conditions} \label{section:initialconditions}

\begin{table*}
	\centering
	\caption{Initial conditions of the star cluster models. }\label{table:details}	
	\resizebox{\textwidth}{!}{\begin{tabular}{llll} 
		\hline\hline
		Model & C12.8k &  C64.8k & C128k\\
		\hline
		Number of stars, $\nstar$ & 12\,800 & 64\,800 & 128\,000 \\
		Number of \MLP{}s, $N_p$ &  12\,800 & 64\,800 & 128\,000 \\
		Stellar initial mass function & \cite{2001MNRAS.322..231K},  $0.08-150~\msun$ & \cite{2001MNRAS.322..231K},  $0.08-150~\msun$ & \cite{2001MNRAS.322..231K},  $0.08-150~\msun$ \\
		MLP mass & Test particles & Test particles & Test particles\\
		Total cluster mass, $M_{\rm cl}$  &   $\sim 7.45\times 10^3\,\msun$ & $\sim 3.7\times 10^4\,\msun$ & $\sim 7.45\times 10^4\,\msun$  \\   
		Dynamical model & \cite{Plummer:1911aa} model & \cite{Plummer:1911aa} model & \cite{Plummer:1911aa} model\\
        \MLP{} spatial distribution & Statistically identical to that of stars & Statistically identical to that of stars & Statistically identical to that of stars \\
        \MLP{} velocity distribution & Statistically identical to that of stars & Statistically identical to that of stars & Statistically identical to that of stars \\
		Half-mass radius, $\halfmass$ & 0.77~pc & 0.77~pc & 0.77~pc\\
		Virial radius & 1 pc & 1 pc & 1 pc \\
		$N$-body (H\'enon) time unit, $T_*$ &  0.18~Myr & 0.08~Myr &  0.06~Myr \\
		Crossing time, $t_{\rm cr}$ &  0.11~Myr & 0.05~Myr &  0.03~Myr  \\
		Half-mass relaxation time, $t_{\rm rh}$ & 27.50~Myr & 51.05~Myr &  66.00~Myr \\
        Stellar evolution & Mass loss enabled & Mass loss enabled & Mass loss enabled \\ 
		External tidal force & Solar neighborhood & Solar neighborhood & Solar neighborhood \\
		Simulation time & 300~Myr & 300~Myr & 300~Myr  \\
		\hline\hline
	\end{tabular}}
\end{table*}

We numerically study the evolution of free-floating \MLP{}s in star clusters by modelling three sets of star cluster simulations, which we will hereafter refer to as models C12.8k, C64.8k, and C128k. The choice of this particular set of models allows us to determine how (i) the different stellar densities affect the dynamical evolution of \MLP{}s, and (ii) which dynamical processes affect the \MLP{}s, and to what degree the \MLP{}s are affected.

The initial conditions for each of the three models are summarized in Table~\ref{table:details}. These models represent star clusters containing $\nstar=12\,800$, $\nstar=64\,800$, and $\nstar=128\,000$ stars respectively. 
The stellar positions and velocities are drawn from a \cite{Plummer:1911aa} distribution in virial equilibrium. All models are initialized with virial radii of $r_{\rm vir}= 1$~pc, corresponding to  initial half-mass radii of $\halfmass \approx 0.77$~pc. Stellar masses are drawn from the \cite{2001MNRAS.322..231K} initial mass function (IMF) in the mass range $0.08-150\,\msun$. The total initial masses of the three clusters are $\clustermass\approx 7.45 \times 10^3 \msun$, $3.7 \times 10^4 \msun$, and $7.45 \times 10^4 \msun$ respectively. The models do not include primordial binaries, and we do not include primordial mass segregation. The initial velocity distributions of the stellar and \MLP{} populations are isotropic. Note that an initially anisotropic velocity distribution strongly affects mass segregation and relaxation in star clusters \citep[e.g.,][]{pavlik2022b,tiongco2022,livernois2022}.
The star clusters evolve in an external tidal field corresponding to that of a star cluster in a Solar orbit in the Milky Way galaxy. The Milky Way is modeled as a point mass of $M_G=9.56\times 10^{10}\,\msun$, with the star cluster in a circular orbit with radius $R_G=8.5$~kpc. The corresponding tidal radius ($r_t$) of a cluster can then be estimated using $r_t\approx(\clustermass/3M_G)^{1/3}R_G$.

In addition to the stars, each model containing $N$ stars also contains $N$ \MLP{}s. The \MLP{}s are assigned positions and velocities with distributions that are statistically identical to those of the stars. Note that in realistic star clusters, ejections of \MLP{}s from their host planetary system may initially lead to a slightly super-virial population of \MLP{}s in the star cluster. However, this population rapidly virializes after a brief epoch of rapid escape at early times; we refer the reader to~\cite{Wang:2015ab} for a discussion on this issue. As the \MLP{}s do not exert a gravitational force on the other bodies, they do not affect the evolution of the stellar population and the other \MLP{}s (see \S~\ref{section:numericalmethod}). 

We evolve all models for 300~Myr, which corresponds to roughly $5t_{\rm rh}$ for model C128k. Model C64.8k is initially $\sim 5$ times denser than model C12.8k, and model C128k is initially $\sim 10$ times denser than model C12.8k. 


\subsection{Timescales of star cluster evolution}\label{section:timescales}

The global evolution of a star cluster in virial equilibrium is governed by its initial crossing time, $\tcross$, its initial half-mass relaxation time, $\trelax$, and its mass segregation timescale, $t_{\rm ms}$. The initial crossing times for the three models are 0.11~Myr, 0.05~Myr and 0.03~Myr, respectively (see Table~\ref{table:details}).  
The half-mass two-body relaxation time \citep{Spitzer:1987aa} is defined as 
\begin{equation}
    \trelax =  
    \frac{0.138\nstar}{\ln\Lambda} 
    \left(
    \frac{\halfmass^3}{G\clustermass}
    \right)^{1/2}
    \quad .
\end{equation}
Here, $\halfmass$ is the half-mass radius, $\ln\Lambda\approx \ln(0.11\nstar)$ is the Coulomb logarithm \citep{Giersz:1994aa}, and $\nstar$ is the number of stars. The mass segregation timescale can be expressed as:
\begin{equation}
    t_{\rm ms} = \frac{\overline{m}}{m_{\rm max}} \trelax \quad ,
\end{equation}
where $\overline{m}$ is the average stellar mass, and $m_{\rm max} \approx 96   \   \msun$ is the mass of the most massive star. The mass segregation timescale is roughly $t_{\rm ms}\approx 0.006\trelax$, resulting in 0.165~Myr, 0.31~Myr and 0.40~Myr for model C12.8k, C64.8k and C128k, respectively.

Note that the properties of the \MLP{} population do not impact $\tcross$, $\trelax$, and $t_{\rm ms}$, as \MLP{}s do not affect the dynamics of the stellar population.


\subsection{Lagrangian radii and mass segregation}\label{section:lagrangian}

\begin{figure}
\centering
\begin{tabular}{c}
  \includegraphics[width=1.0\columnwidth]{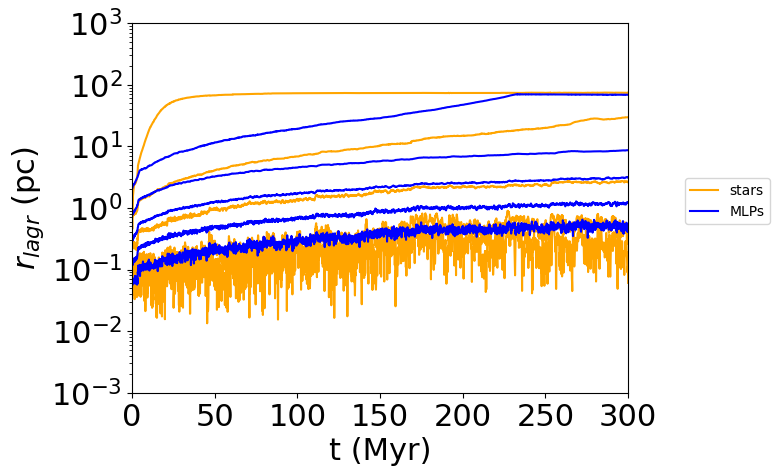} \\  
  \includegraphics[width=1.0\columnwidth]{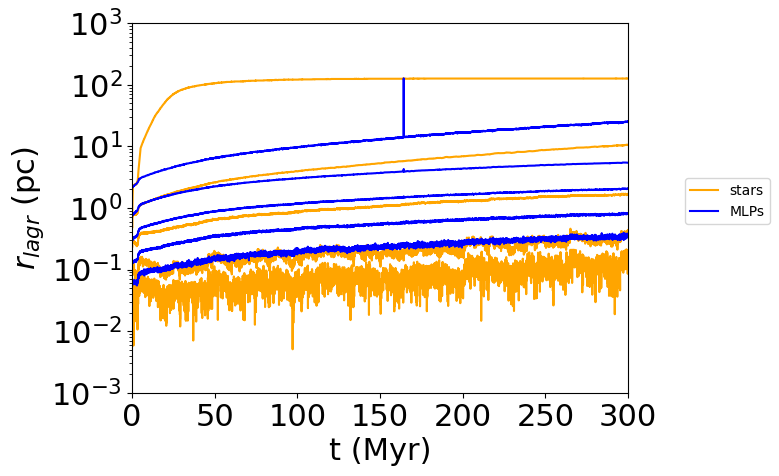}\\  
  \includegraphics[width=1.0\columnwidth]{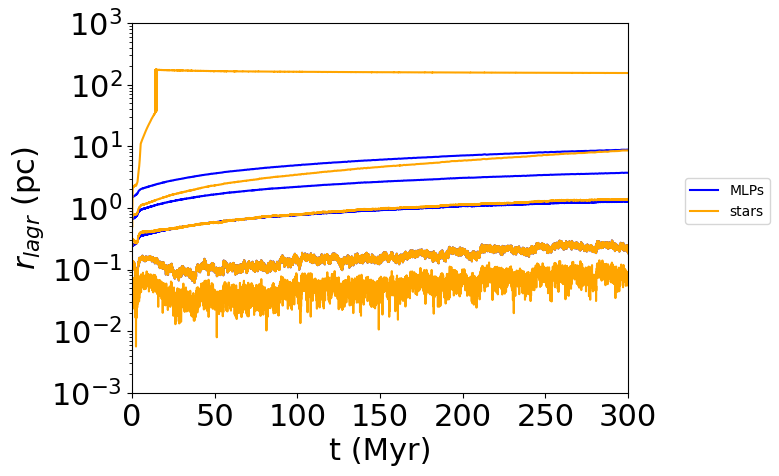}
\end{tabular}
 \caption{Evolution of the Lagrangian radii containing 0.1 \% (\textit{bottom}), 1\%, 10\%, 50\% and 90 \% (\textit{top}) of the cluster mass, for stars (in green) and \MLP{}s (in red) for model C12.8k (top panel), C64.8k (middle panel) and C128k (bottom panel).  The Lagrangian radii at each time are calculated using the total cluster mass of the star cluster at that time.
 } \label{fig:lagrComparing}
\end{figure}

\begin{figure}[h!]
\begin{tabular}{c}
  \includegraphics[width=0.99\columnwidth]{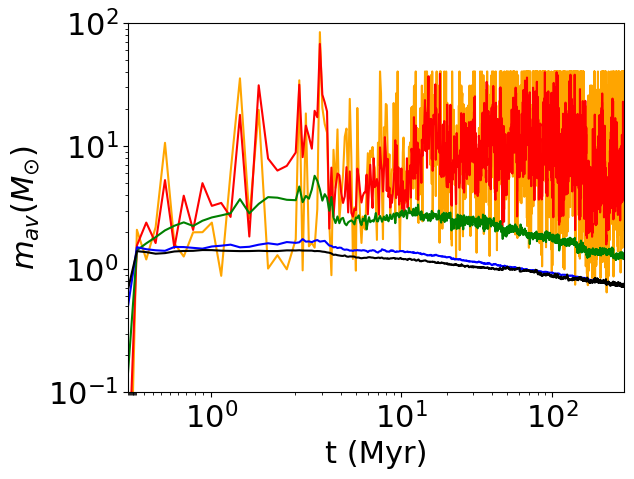}\\
    \includegraphics[width=0.99\columnwidth]{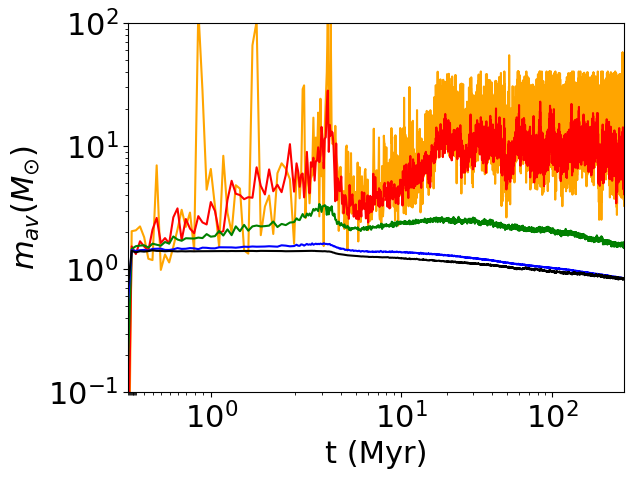}\\
  \includegraphics[width=0.99\columnwidth]{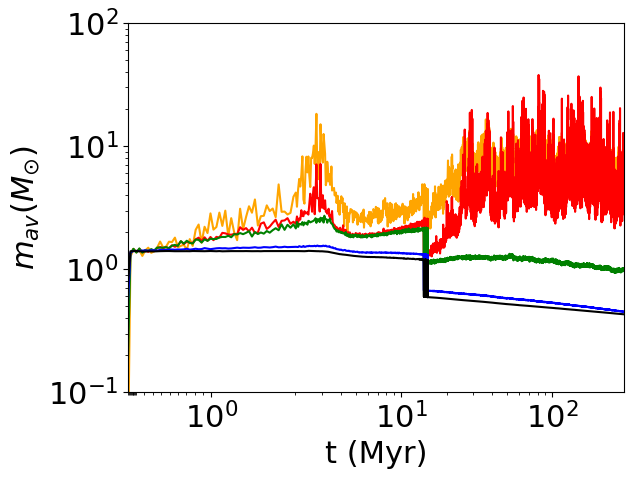}
\end{tabular}
  \caption{The evolution of average stellar mass within the 0.1 \% ({\it orange curve}), 1 \% ({\it red curve}), 10 \% ({\it green curve}), 50 \% ({\it blue curve}) and 90 \% ({\it black curve}) Lagrangian shells, for models C12.8k (top), C64.8k (middle) and C128k (bottom). The evolution in the first Myrs is dominated by stellar evolution and mass segregation, and it is more steady in all shells around 20 Myrs. The average mass of the inner Lagrangian shells changes more frequently due to the lower number of stars in the shells. This is more evident in the low density model.  
  }\label{fig:avmass}
\end{figure}

Lagrangian radii provide a powerful tool for analysing the global evolution of a star cluster, as well as for characterizing mass segregation. The evolution of the Lagrangian radii for the three models is shown in Figure~\ref{fig:lagrComparing}, for Lagrangian radii ranging from the 0.1 \% mass shell to the 90\% mass shell.\\
The Lagrangian radii for the stellar population are calculated using the total cluster mass at each time. The Lagrangian radii for the population of \MLP{}s are calculated from the number of \MLP{}s enclosed in the Lagrangian shell at that time (instead of the mass, since \MLP{}s are massless). 

During the core collapse phase of a star cluster, strong gravitational scattering events result in the ejection of stars to (and beyond) the outskirts of the star cluster. Simultaneously, stellar evolution reduces the total mass of the star cluster. As a result of stellar ejections and stellar mass loss, the gravitational potential of the star cluster gradually decreases, and \MLP{}s slowly migrate towards the outskirts of the star cluster. This results in a divergence between the Lagrangian radii for the stars and the Lagrangian radii for the \MLP{}s. As the expansion of the \MLP{} population follows the gravitational potential of the star cluster, the Lagrangian radii for the \MLP{}s are typically larger than the corresponding Lagrangian radii of the stellar population, except in the outskirts of the cluster, which is dominated by stars ejected from the core, primarily during the core collapse phase. 

The evolution of the Lagrangian radii is similar in all models, but there are notable variations due to the differences in stellar number density of the star clusters. The inner regions display a significant change between the dynamical evolution of both stars and \MLP{}s. The innermost regions in Figure~\ref{fig:lagrComparing} show a less pronounced core collapse for the Lagrangian radii of the \MLP{}s, when compared to the Lagrangian radii of the stars. The expansion of the population of \MLP{}s in the inner regions is less pronounced when the stellar density is higher. In model C128k, for example, the 10 \% shell of both \MLP{}s and stars is similar.\\
The 50 \% Lagrangian radii of the stars and \MLP{}s evolve similarly, but show a divergence after the core collapse. The stellar 50 \% Lagrangian radius grows faster than the corresponding \MLP{} Lagrangian radius in model C12.8k.\\
The stellar Lagrangian 90 \% shell is particularly interesting. 
In dense models, after the initial core collapse, there is a larger ejection of mass in the cluster. The former leads to a significant expansion of the external regions due to the ejection of high-energy stars, particularly black holes and neutron stars, which is due to the stellar evolution in the first Myrs. The large expansion of the shell can be explained by the relatively mild collapse of the core in the inner regions. This causes high-energy stars to be ejected, leading to an earlier cooling of the core. The 90 \% Lagrangian shells of the \MLP{}s follow a similar evolution in all models, regardless of the density of the star cluster. In particular, the 50 \% Lagrangian shell of the stars in model C128k becomes similar to the 90 \% Lagrangian shell of the \MLP{}s, indicating that \MLP{}s are more likely to remain bounded to the star cluster. 
The spike in the 90 \% shell of the \MLP{}s Lagrangian radius in the middle panel of Figure~\ref{fig:lagrComparing} is causd by an \MLP{} that is ejected from the star cluster at the next timestep. 
More massive inner regions are more likely to bound the \MLP{}s. Thus they migrate much slower to the outer regions than the stars. The behaviour of the 90 \% shell is related to (i) the ejection of a large number of hard binaries, which make a large mass contribution to outer shells and (ii) black holes and neutron stars, which contribute also to a less strong core collapse in the denser models.

In stellar systems containing two populations that have considerable differences in particle masses, the high-mass component tends to decouple from the low-mass component within a short period of time \citep[see][]{Khalisi:2007aa}. As the \MLP{}s are massless, they do not participate in the energy equipartition process that is responsible for mass segregation of the stellar population. Free-floating \MLP{}s thus have relatively unperturbed trajectories, and generally follow the overall gravitational potential of the star cluster. 

Close encounters between stars and \MLP{}s do occur \citep[see, e.g.,][]{Wang:2015aa}. During a two-body scattering event between a star and a \MLP{}, the motion of the star remains unperturbed. In the absence of other massive perturbers nearby, the \MLP{}'s trajectory in the two-body encounter can be approximated with an hyperbolic orbit, during which the direction of its velocity vector changes, while the magnitude of the velocity vector remains constant (when evaluated at times sufficiently long before and after the encounter). The latter interaction therefore generally does not result in the ejection of a \MLP{}, as its speed remains below the local escape velocity. 

Close encounters involving three stars can result in the exchange of a substantial amount of kinetic energy, and may lead to the ejection of one of the stars, usually the lowest-mass star. Three-body encounters involving two stars and one \MLP{} occur less frequently, as the trajectories are less affected gravitational focusing (with respect to an encounter between three stars). When the three bodies approach each other at sufficiently close distances, this process contributes to the ejection of \MLP{}s following gravitational slingshot events \citep[see, e.g.,][]{Saslaw:1974aa, Wang:2015aa}. Three-body encounters involving one star and two \MLP{}s can be treated as two separate two-body encounters, as the two \MLP{}s do not interact with each other, and therefore do not lead to ejection of any of the bodies from the system.


\subsection{Mass segregation}

The presence of a mass spectrum in the stellar population results in rapid mass segregation (see Figure~\ref{fig:avmass}). This process is a consequence of the tendency of the stellar population to achieve local energy equipartition, although the stellar population is unable to achieve energy equipartition through this process \citep[see][]{parker2016}. For details on the mass segregation process itself, we refer to the studies of~\cite{Khalisi:2007aa} and~\cite{Allison:2009aa}.

Figure~\ref{fig:avmass} shows the average stellar mass contained within several Lagrangian shells. The scatter in the curves representing the innermost Lagrangian radii is affected by small-number statistics. During the first million years, the tendency of stars to achieve local energy equipartition causes the most massive stars to sink to the centre, where they experience close encounters \citep[e.g.,][]{Khalisi:2007aa}. This causes rapid variations in the average mass in the core. These strong fluctuations are gradually mediated by stellar evolution and expansion of the core. Throughout most of the simulations, the average mass in the core region is typically a factor ten higher than that in the outskirts of the star cluster. 

The average stellar mass within the 1 \% Lagrangian radius (blue curves in Figure~\ref{fig:avmass}) increases substantially as the cluster evolves. The initial increase ($t<5$~Myr) is due to mass segregation, and the substantial decrease at later times is caused by stellar evolution. Inspection of the output files of the simulations indicates that the central region is initially dominated by several massive stars. As most massive stars have evolved at times $t\ga 10$~Myr, stellar evolution becomes less prominent, and the average mass in the core is dominated by stellar mass segregation. In the denser star cluster models, mass segregation occurs at later times, mass segregation is more prominent, and the outer regions abruptly stabilise at the time of the core collapse.


\subsection{Escaping stars and \MLP{}s}\label{section:escapingstar}

\begin{figure*}[h!]
\begin{tabular}{c}
  \includegraphics[width=1.0\columnwidth]{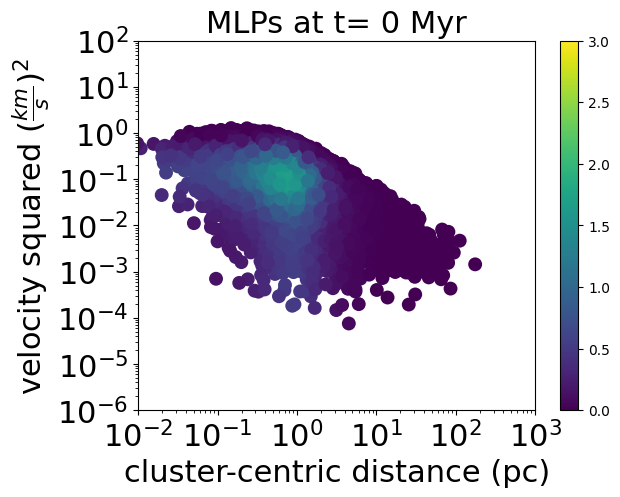} \\
  \includegraphics[width=\columnwidth]{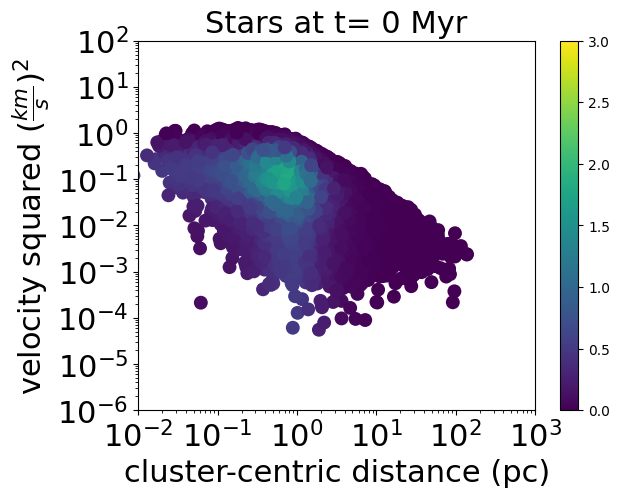}
    \includegraphics[width=\columnwidth]{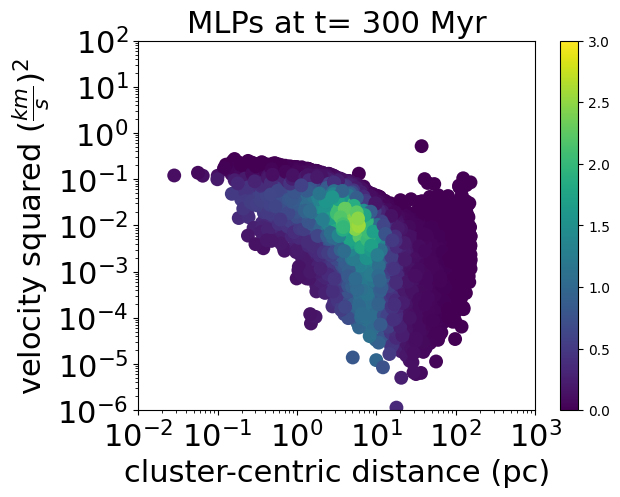} \\
  \includegraphics[width=\columnwidth]{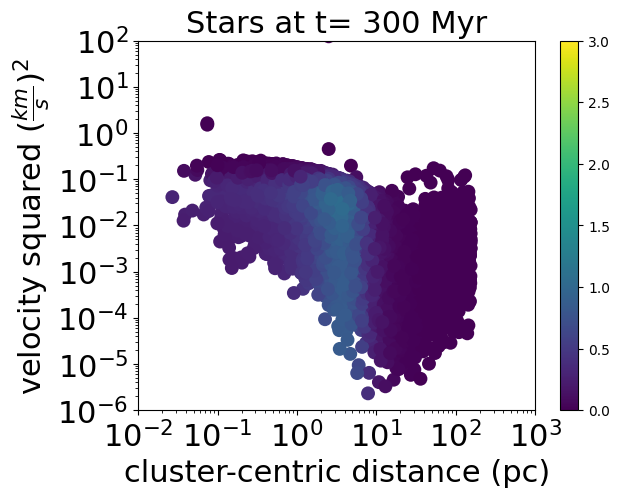}
\end{tabular}
  \caption{Distances and speeds of \MLP{}s (left column) and stars (right column) at $t=0$~Myr ({\it top}) and at $t=300$~Myr ({\it bottom}) for model C128k. Colors indicate the number density of the data in the plot, as shown in the colorbar. For the methodology used, we refer to \protect\cite{flammini2023}.
  }\label{fig:rVSv2_70000}
\end{figure*}

\begin{figure}
\begin{tabular}{c}
  \includegraphics[width=1.0\columnwidth]{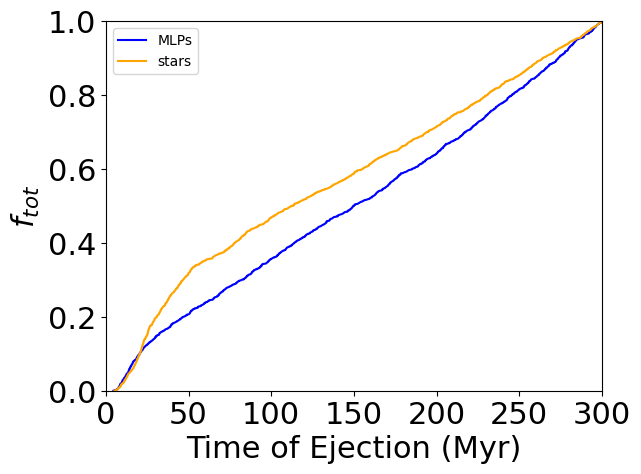}\\
  \includegraphics[width=1.0\columnwidth]{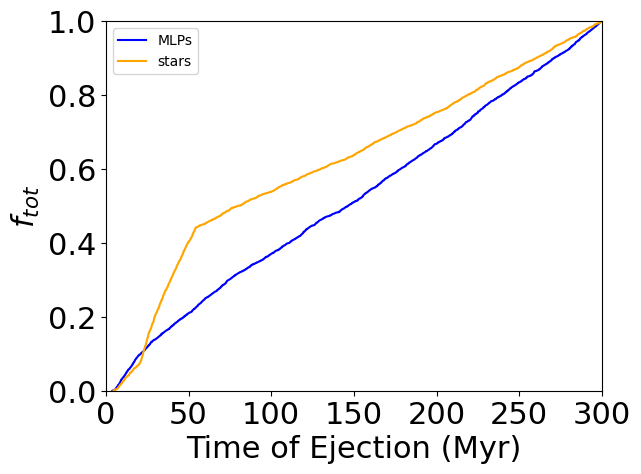}\\
  \includegraphics[width=1.0\columnwidth]{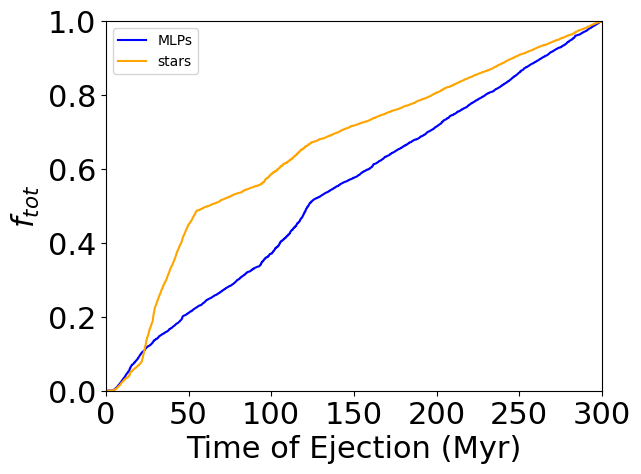}
\end{tabular}
  \caption{The cumulative distribution of ejection times for \MLP{}s (blue) and stars (orange) for models C12.8k (top), C64.8k (center) and C128k (bottom).}\label{fig:cumdistrejec}
\end{figure}

\begin{figure}
\begin{tabular}{c}
  \includegraphics[width=1.0\columnwidth]{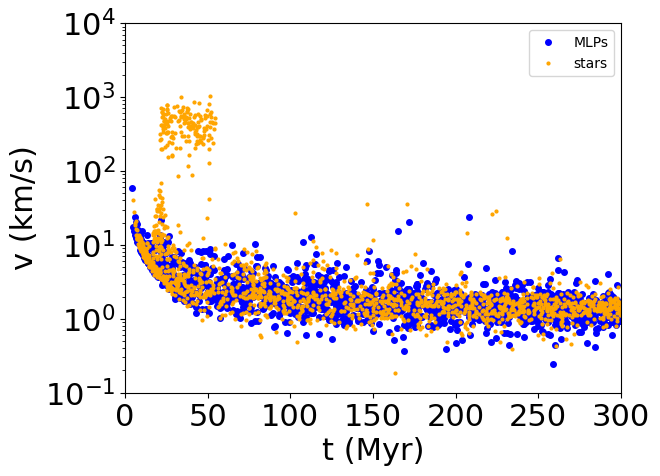}\\
  \includegraphics[width=1.0\columnwidth]{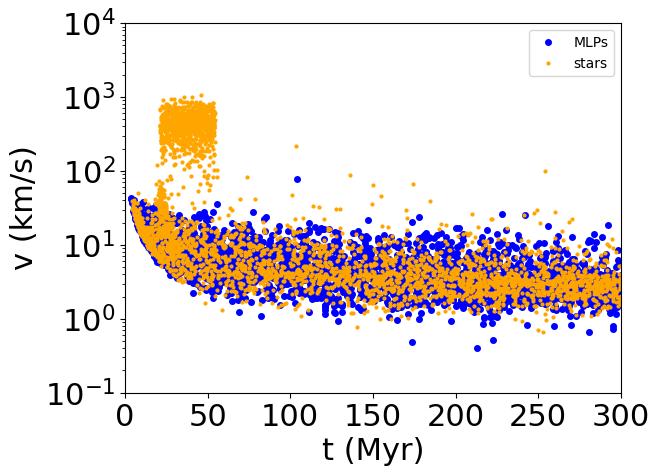}\\
  \includegraphics[width=1.0\columnwidth]{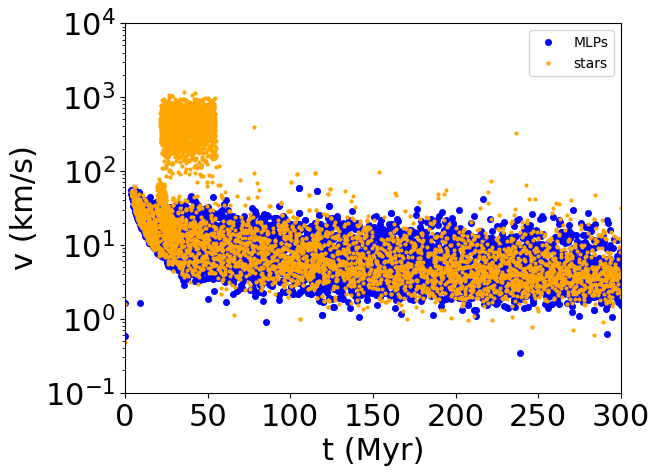}
\end{tabular}
  \caption{The ejection velocity distribution of \MLP{}s (blue) and stars (red) for models C12.8k (top), C64.8k (center) and C128k (bottom). 
  }\label{fig:escapeveldist}
\end{figure}

\begin{figure}
  \includegraphics[width=1.0\columnwidth]{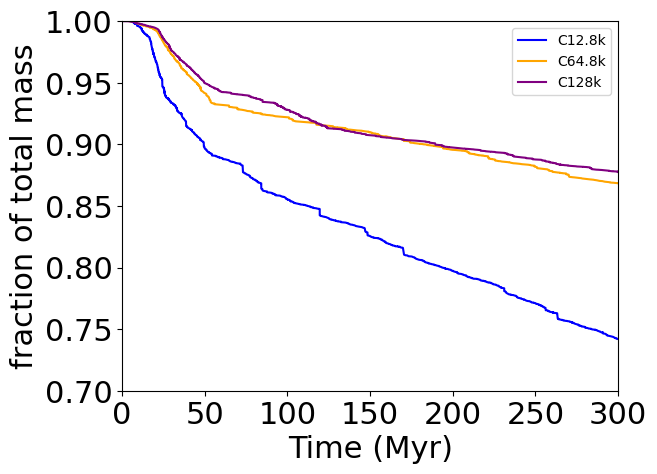}\\
  \caption{Evolution of the total mass of the star cluster, for all models. The total mass of the cluster gradually decreases over time due to stellar evolution and due to escaping stars. Note that the total mass is independent of the number of \MLP{}s. 
  }\label{fig:massEvolution}
\end{figure}

Figure~\ref{fig:rVSv2_70000} shows the radial distances and speeds of all stars and \MLP{}s in the C128k model, at times $t=0$~Myr and $t=300$~Myr. Stars are ejected from the core region of each cluster, as a result of close encounters with other stars, and mass is lost due to stellar evolution. Most of the stars in the outskirts of the star cluster migrate to outer regions as a consequence of scattering with other stars, and to a lesser extent due to the reduced gravitational potential of the star cluster following stellar evolution and stellar escapers. Most of the \MLP{}s that migrate to the outskirts, do so because of the decreasing gravitational potential of the star cluster, and tend to escape over time, as the tidal radius of the star cluster shrinks while star cluster experiences mass loss due to stellar evolution and stellar escapers. 
%

Figure~\ref{fig:cumdistrejec} shows the cumulative number of escaping stars and \MLP{}s over time. The distribution of escape speeds over time is shown in Figure~\ref{fig:escapeveldist}, and the evolution of the total cluster mass is shown in Figure~\ref{fig:massEvolution}.\\
The escape rate of stars is higher during the initial phases of evolution due to the initial core collapse, and it  gradually decreases as the cluster expands. The effect is definitely more prominent and lasts longer in dense star clusters. The subsequent expansion of the core, combined with the total mass, results in a longer relaxation time, implies that the two-body relaxation in the core, and consequently the stellar ejection rate, is suppressed \citep[see][]{Kouwenhoven:2014aa}. 
Therefore, stars are ejected mostly due to the core collapse at the start of the simulation, and due to subsequent mass segregation. During the first 50~Myr, the ejection of high-energy stars (Figure~\ref{fig:escapeveldist}), is more prominent in denser star clusters. Most are neutron stars and black holes, which also indicate the prominent role of stellar evolution on the evolution of the star cluster.
The average speed of the ejected stars is 43.91\,\kms, 107.14\,\kms and 134.56\,\kms in models C12.8k, C64.8k, C128k, respectively. The average speed of the ejected \MLP{}s is 2.83\,\kms, 6.52 \kms and 9.42\,\kms for models C12.8k, C64.8k, C128k respectively. The majority of the stellar escape events occur during the initial phases of the star cluster evolution. During this period, there are a large number of high-velocity escapers, which significantly impact the average escape speed. In denser models, this region becomes even more populated. If we exclude these black holes and neutron stars from the average stellar escape speed, we obtain an average escape speeds of 3.86\,\kms, 8.26\,\kms and 11.56\,\kms for models C12.8k, C64.8k, C128k, respectively. The escaping \MLP{}s thus have similar, but slightly lower, average escape speeds, compared to the stellar escapers.

We do not find escaping \MLP{}s with speeds $>10^2$\,\kms. The majority of stars with such high velocity are either binary systems or compact objects, and are therefore unlikely to be found for a \MLP{}. The zero mass of an \MLP{} reflects in the total energy of the particle, which can only be enhanced to these velocities by interactions with stars or, more effectively, by compact objects. 
To summarise, stars acquire large velocities also as a consequence of mass segregation and stellar evolution. Instead, planets acquire large velocities mostly with the interaction with stars. Figure~\ref{fig:escapeveldist} shows that the escape speed distribution is mostly similar for stars and \MLP{}s, with a somewhat broader escape speed distribution for the \MLP{}s. 
After the first 50~Myr, the number of ejected stars is roughly 37\%, 42\% and 50\% of the total number of ejected stars at the end of the simulation. Therefore, these stars have a larger statistical weight on the average velocity.
The stars are therefore mainly ejected from the star cluster, in the first 50 Myrs, due to the initial core collapse and stellar evolution. After the first 50 Myrs, the ejection of stars from the star cluster is due to either evaporation or ejection. 
The \MLP{}, instead, are ejected from the star cluster by either evaporation or ejection, and they are not instead related to the mechanism of mass segregation. This is another important result, as it show there is no mass segregation impact on planet-mass objects.

Finally, we show the evolution of the total mass of each cluster in  Figure~\ref{fig:massEvolution}. During the 300~Myr of evolution, cluster models C12.8k, C64.8k and C128k lose 25.81\%, 13.16\% and 12.25\% of their initial mass, respectively. The majority of mass is lost before 50~Myr for all models. Thus, stellar evolution greatly impact the mass loss in denser star cluster models, which strongly reduce the mass loss. The mass loss is roughly linear with evaporation from the star cluster and high-energy stars for the remainder of the simulations. To summarise, the ejection of high energy stars impact the loss of mass, while this contribution become less important for the remainder of the simulation. Moreover, denser star cluster models retain more mass due to their deeper gravitational potential wells.


\section{Conclusions}\label{section:conclusions}

We present the first work done with the newly-developed code \nbm{}, which is based on the direct $N$-body code \nb{}. \nbm{} is optimised for integrating star clusters with a large number of massless particles (\MLP{}s), such as planets, comets, asteroids, and planetesimals. Using theoretical arguments and numerical performance tests, we demonstrate that \nbm{} outperforms \nb{} significantly, when a large number of \MLP{}s is present. Through simulations of different star cluster models with a large number of \MLP{}s, we analyse the dynamical evolution of both the stellar and \MLP{} populations, with an emphasis on the evolution of the Lagrangian radii, mass segregation, and the properties of escaping particles. Our main results are summarised as follows:
\begin{enumerate}

    \item We have developed the code \nbm{}, a modified version of \nb{}, which allows fast evolution of star clusters that contain a large number of \MLP{}s. This is achieved through excluding the gravitational forces of \MLP{}s on any of the other particles. We show that, even when the number of \MLP{}s is comparable to the number of stars, the wall-clock time is only moderately longer compared to the simulation without \MLP{}s.
    
    \item \MLP{}s do not participate in the mass segregation process, as they are unable to change energy with the stellar population as they are test particles. In terms of their dynamical evolution, our study finds that: (i) although the spatial distributions of the stellar population and the \MLP{} population diverge over time, this process should not be confused with the classical mass segregation process for the stellar population. Unlike the stellar population, the kinematic evolution of a population of \MLP{}s in a star cluster is primarily determined by the overall changes in the gravitational potential of the cluster, which is reduced over time due to stellar ejections and stellar evolution. 
    This occurs when modeling \MLP{}s as massless, and also when assigning them a small mass. Thus, modeling \MLP{}s as massless is an adequate approximation (see Figure~\ref{fig:standardvsmassless}).

    \item The dynamical evolution of the stellar population in star clusters with a higher initial stellar density is more strongly dominated by the initial core collapse, stellar evolution and mass segregation, especially during the first 50~Myr. The population of \MLP{}s, on the other hand, evolves similarly in all star clusters models. 
    
    \item The cumulative number of escaping \MLP{}s over time is smoother compared to that of the stellar escapers. There are two main reason for this behavior: (i) stellar evolution, which is especially represented by the high energy stars ejected during the first 50 Myr, mainly neutron stars, and (ii) mass segregation, which inevitably cause ejection of stars in the outer regions. This also tell us that those particles are mainly ejected through evaporation or ejection, but these processes are not influenced by mass segregation and stellar evolution, as stars. 
    
    \item The average escape speeds of stars and \MLP{}s are different.
    This is caused mainly by (i) stellar evolution, where high-velocity stars, such as neutron stars, are ejected, and (ii) density dependence, as the escape velocity is higher for denser star clusters. If we do not consider these high energy stars during the first 50~Myrs (i.e., we if ignore the contribution of stellar evolution to the stellar escape rate) then the velocities of ejection of star and \MLP{} are comparable. This also reflects what occurs during the Lagrangian evolution of the particles at the beginning of the simulation.
 
\end{enumerate}

The parameter space covered by the models in our study is limited, but provides useful insights about the differences between the dynamical evolution of the stellar population and the \MLP{} population in a star cluster. We have not included primordial (stellar) binary systems in our simulations. 
In our current study, we have also not included any planetary companions among star cluster members. Both these feature will be added in the subsequent papers in this series.

\begin{acknowledgements}

We are grateful to Long Wang for his support during the initial phase of this project.

FFD and RS acknowledge support by the German Science Foundation (DFG), priority program SPP 1992 "Exploring the diversity of extrasolar planets" (project Sp 345/22-1 and central visitor program).

MBNK was supported by the National Natural Science Foundation of China (grant 11573004).

FFD. and MBNK were supported by the Research Development Fund (grant RDF-16–01–16) of XJTLU. FFD acknowledges support from the XJTLU postgraduate research scholarship. This publication was made possible through the support of a grant from  the John Templeton Foundation and National Astronomical Observatories of the  Chinese Academy of Sciences. 

PB thanks the support from the special program of the Polish Academy of Sciences and the U.S. National Academy of Sciences under the Long-term program to support Ukrainian research teams grant No.~PAN.BFB.S.BWZ.329.022.2023.

PB acknowledges the support by the National Science Foundation of China under grant NSFC~No.~12473017.

\end{acknowledgements}

%
%

\bibliographystyle{aa}
\bibliography{aanda}

\end{document}